\documentclass[runningheads]{llncs}
\usepackage[T1]{fontenc}
\usepackage{graphicx,verbatim}
\usepackage{multirow} 
\usepackage{amssymb} 
\usepackage{pifont}
\usepackage{makecell}
\usepackage{adjustbox} 
\usepackage{hyperref}
\usepackage{amsmath} 
\usepackage{float} 
\usepackage{subcaption}
\usepackage{booktabs}
\begin{document}
\title{VMRA-MaR: An Asymmetry-Aware Temporal Framework for Longitudinal Breast Cancer Risk Prediction}
\author{Zijun Sun\inst{1} \and
Solveig Thrun\inst{2} \and
Michael Kampffmeyer\inst{2,3}}
\authorrunning{Z. Sun et al.}
%
\institute{Department of Computer Science and Engineering, University of Bologna, \\
Bologna, Italy
\email{zijun.sun@studio.unibo.it}
\and Department of Physics and Technology, UiT The Arctic University of Norway, Tromsø, Norway\\
\and Norwegian Computing Center, Oslo, Norway
}

\titlerunning{Asymmetry-Aware Temporal
Framework for Breast Cancer Risk
Prediction}
    
\maketitle              

\begin{abstract}
Breast cancer remains a leading cause of mortality worldwide and is typically detected via screening programs where healthy people are invited in regular intervals. Automated risk prediction approaches have the potential to improve this process by facilitating dynamically screening of high-risk groups. While most models focus solely on the most recent screening, there is growing interest in exploiting temporal information to capture evolving trends in breast tissue, as inspired by clinical practice. Early methods typically relied on two time steps, and although recent efforts have extended this to multiple time steps using Transformer architectures, challenges remain in fully harnessing the rich temporal dynamics inherent in longitudinal imaging data. In this work, we propose to instead leverage Vision Mamba RNN (VMRNN) with a state-space model (SSM) and LSTM-like memory mechanisms to effectively capture nuanced trends in breast tissue evolution. To further enhance our approach, we incorporate an asymmetry module that utilizes a Spatial Asymmetry Detector (SAD) and Longitudinal Asymmetry Tracker (LAT) to identify clinically relevant bilateral differences. This integrated framework demonstrates notable improvements in predicting cancer onset, especially for the more challenging high-density breast cases and achieves superior performance at extended time points (years four and five), highlighting its potential to advance early breast cancer recognition and enable more personalized screening strategies. Our code is available at \href{https://github.com/Mortal-Suen/VMRA-MaR.git}{this URL}.


\end{abstract}

\section{Introduction}
Breast cancer remains the most prevalent malignancy worldwide, imposing a significant public health burden \cite{bray2024global,hakama2008cancer}. Early detection through precise risk prediction is pivotal for improving clinical outcomes and reducing mortality. Despite debates over screening intervals, optimal starting age, and the utility of supplemental imaging, digital mammography remains the primary screening modality due to its cost-effectiveness and broad accessibility \cite{wilkinson2025cost,gross2013cost}. These ongoing controversies underscore the pressing need for enhanced risk assessment models that enable personalized screening regimens while minimizing unnecessary interventions \cite{otto2014controversies}.\\
\indent While traditional risk models \cite{evans2007breast,terry201910,Tyrer-Cuzick} have achieved only moderate predictive accuracy, deep learning–based methods have substantially advanced breast cancer risk stratification \cite{dl_in_breast_cancer}. A seminal model is Mirai \cite{Mirai}, which achieves strong performance by integrating risk factor data, managing missing information, and ensuring consistency across different mammography machines. Nonetheless, its primary focus on the most recent mammogram overlooks the abundant temporal information in sequential mammograms, an aspect that could reveal subtle indicators of disease progression and further refine early detection strategies.
\indent There has been a growing interest in addressing this shortcoming by incorporating longitudinal imaging data. While early methods typically rely on two or three years of patient history \cite{RADIFUSION,PRIME+}, recent work such as LoMaR \cite{LoMaR} expands upon this approach using a hybrid CNN–Transformer~\cite{Transformer} architecture that leverages more extensive imaging history for more accurate risk predictions. Despite demonstrating the benefits of comprehensive longitudinal data, LoMaR and similar models often depend on positional embeddings, which may underutilize the full temporal richness available in large-scale clinical datasets.\\
\indent In this study, we introduce VMRA-MaR, a novel recurrent-based framework designed to address limitations in modeling dynamic breast tissue changes over time. Drawing inspiration from AsymMirai~\cite{AsymMirai}, we further incorporate bilateral symmetry analysis into the risk prediction process, thereby accounting for contralateral asymmetry within an end-to-end learning pipeline. The model is validated using the CSAW-CC dataset from Karolinska University Hospital, which provides comprehensive longitudinal mammographic data. Our primary contributions are threefold: a) We propose a VMRNN-based~\cite{VMRNN} framework that leverages time-step information to track and model longitudinal breast tissue changes more effectively than models reliant on positional embeddings alone; b) We incorporate asymmetry features into risk prediction, building on AsymMirai’s insight that contralateral asymmetry strongly correlates with future cancer risk, while extending its application to a longitudinal context; c) By utilizing multi-year sequential mammography data, our approach improves risk prediction accuracy and supports earlier detection of subtle disease progression, particularly in high-density cases.

\section{Method}
\subsection{Overall Pipeline}
\begin{figure}[tp]
    \centering
    \includegraphics[width=1\linewidth]{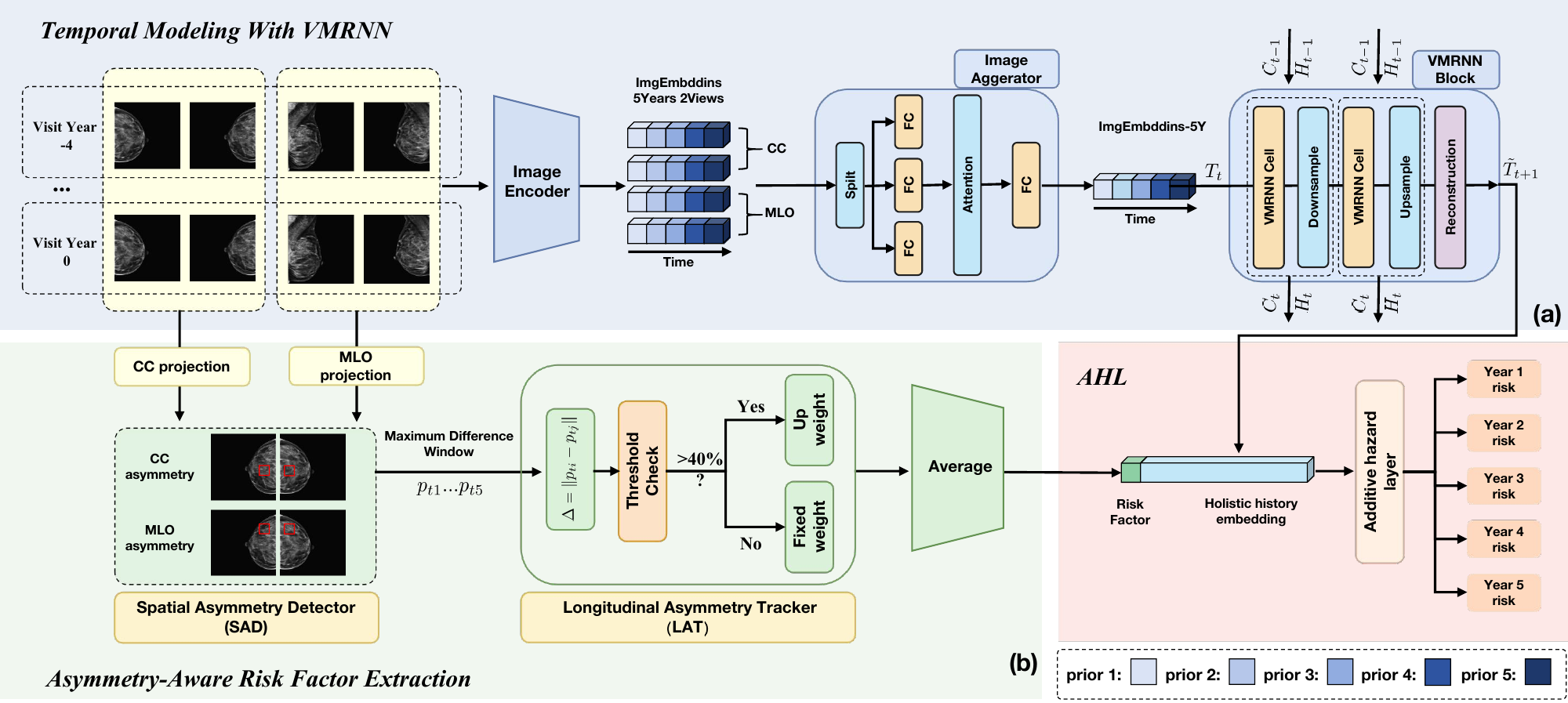}
    \caption{Network architecture of our proposed VMRNN-Asymmetry Mammogram Risk model (VMRA-MaR).}
    \label{fig:VMRA-MaR}
\end{figure}

Figure~\ref{fig:VMRA-MaR} provides an overview of our unified approach, which integrates and extends existing spatiotemporal and asymmetry-detection techniques to address limitations of prior models that underutilize temporal features and breast symmetry. Specifically, we fuse the temporal representations from the Vision Mamba RNN (VMRNN), with asymmetry cues within an Additive Hazard Layer to generate a five-year risk prediction. This multisource information fusion strategy more effectively captures evolving breast tissue characteristics, identifies persistent asymmetry signals, and offers a comprehensive view of each patient’s risk trajectory.

Our model processes five yearly screenings, each comprising four standard mammographic views (LCC, RCC, LMLO, RMLO). At each time step $t$ and viewpoint $v$, the corresponding image $x_t^v$ is first encoded via a pre-trained image encoder; the resulting multi-view features are then fused using a Transformer module. These fused features are passed to the VMRNN, which learns longitudinal dependencies through the evolving hidden states $\left\{H^1, \ldots, H^T\right\}$. Concurrently, a Spatial Asymmetry Detector (SAD) identifies regional discrepancies between left-right views, and a Longitudinal Asymmetry Tracker (LAT) aggregates these differences over time to highlight persistent anomalies. Finally, the AHL combines both the temporal representations and asymmetry cues to produce a unified five-year risk prediction that accounts for evolving breast tissue features as well as persistent asymmetry signals.

\subsection{Temporal Modeling within VMRA-MaR}
The current state-of-the-art (SOTA) approach to leverage longitudinal imaging data is LoMaR~\cite{LoMaR}, which leverages a Transformer backbone and positional embeddings to integrate the temporal information. However, this design relies heavily on the global attention mechanism of the transformer and its static positional embeddings, struggling to fully capture temporal coherent features. VMRA-MaR instead builds on a VMRNN block~\cite{VMRNN}, that processes data recurrently updating hidden states at each time step, allowing us to address this shortcoming, resulting in improved accuracy when modeling longitudinal changes. 

\subsubsection{VMRNN Block} 
The VMRNN Block (see Figure~\ref{fig:VMRA-MaR}(a)) processes, at each time step $t$, a fused feature vector $T_t$ (obtained from the Transformer Block shown in Figure~\ref{fig:VMRA-MaR}(a)), along with the previous hidden state $H_{t-1}$ and cell state $C_{t-1}$. The block incorporates downsampling and upsampling layers to capture multi-scale representations, and a reconstruction layer to restore spatial resolution for subsequent predictions. 

Within the block, the VMRNN Cell first projects $T_t$ and $H_{t-1}$ into an intermediate representation using a linear projection (LP) layer that fuses and reshapes the features into a spatial format suitable for further processing by a VSS Block:
\begin{equation}
    X_t=\mathrm{LP}\left(T_t, H_{t-1}\right)
\end{equation}
The VSS refines $X_t$ into the gating signal $F_t$ by fusing directional and direct spatial features. First, $X_t$ is reshaped into an image-like tensor $A$. In the primary stream, a depth-wise $3 \times 3$ convolution followed by a SiLU activation produces.
\begin{equation}
   A_1=\operatorname{SiLU}\left(\mathrm{DWConv}_{3 \times 3}(A)\right) 
\end{equation}
which is expanded into four directional sequences $\left\{A_{1, v}\right\}_{v=1}^4$. Each sequence is processed by a directional module $S 6$~\cite{Mamba}, and the resulting directional features $\left\{\bar{A}_v\right\}$ are merged and normalized:
\begin{equation}
    \bar{A}_v=S 6\left(\operatorname{expand}\left(A_1, v\right)\right), \quad A_3=\operatorname{LayerNorm}\left(\operatorname{merge}\left(\bar{A}_1, \bar{A}_2, \bar{A}_3, \bar{A}_4\right)\right)
\end{equation}
In parallel, a secondary stream applies a SiLU activation~\cite{silu} to $A$. The two streams are combined by element-wise addition,
\begin{equation}
    B_1=\operatorname{SiLU}(A),\ Y=A_3+B_1
\end{equation}
and a sigmoid activation is then applied to produce $F_t=\sigma(Y)$. By fusing directional and direct spatial features, the VSS provides a context-aware gating signal that enhances the model's ability to capture fine-grained spatial dependencies across screening time steps.

Subsequently, $Y$ is used to derive a gating signal $F_t$ via a sigmoid $\sigma(Y)$ and the cell and hidden states are updated as follows: 

\begin{equation}
    C_t=F_t \odot\left(\tanh (Y)+C_{t-1}\right),\ H_t=F_t \odot \tanh \left(C_t\right)
\end{equation}

\subsection{Asymmetry-Aware Risk Factor Extraction}
Building on the insight of AsymMirai~\cite{AsymMirai} that asymmetry is a pivotal marker for risk prediction in single time‐step mammographic screenings, we incorporate asymmetry into our end-to-end framework and extend their approach to further consider the longitudinal dynamics inherent in mammographic screenings. Our novel VMRNN-based framework leverages a Spatial Asymmetry Detector (SAD) to capture interbreast differences and a Longitudinal Asymmetry Tracker (LAT) to integrate temporal variations, thereby enhancing overall risk evaluation.

In a single screening, following~\cite{AsymMirai}, SAD aligns the left and right feature maps $L$ and $R$ by flipping $R$ to obtain $R^F$, and then computes the local feature differences $D$ to determine the maximum asymmetry value $D_{\max }$ along with its spatial coordinates $\mathbf{p}$. 
\begin{equation}
    \begin{split}
        D_{\mathrm{norm}}(h, w)=\sqrt{\sum_{c=1}^C D_{c, h, w}^2} ,\ D_{\max }=\max _{h, w}D_{\mathrm{norm}}(h, w),
    \end{split}
\end{equation}
where $\mathbf{p}$ denotes the coordinates at which $D_{\max }$ is attained. Building on the single time-step asymmetry, our VMRA-MaR consists of the LAT to analyze the Euclidean displacement of $\mathbf{p}$ across screenings; if the displacement is below a predefined threshold ($40 \%$ of the window size), the region is considered persistently abnormal, and its risk score is upweighted via weighted fusion of historical measurements.

\subsection{Additive Hazard Layer (AHL)}
The additive-hazard layer (AHL)~\cite{AHL} integrates the entire screening history embedding, $\mathcal{H}$, with an asymmetry-aware risk factor, $r_{A A}$, extracted from the dedicated module. These components are concatenated to form $\widetilde{R}=$ concat $\left[\mathcal{H}, r_{A A}\right]$, which serves as the input for subsequent risk estimations. A baseline risk, $B(\widetilde{R})$, is computed via a ${Linear}_k(\widetilde{R})$. In parallel, timespecific marginal hazards are estimated through separate linear layers followed by ReLU activations, thereby ensuring non-negative hazard increments and enforcing the monotonicity of cumulative risk predictions. Formally, for each time interval $K$, the marginal hazard is defined as:
\begin{equation}
H_k(\widetilde{R})=\operatorname{ReLU}\left(\operatorname{Linear}_k(\widetilde{R})\right) 
\end{equation}
and the cumulative risk of developing cancer by time $k$ is given by 
\begin{equation}
    P\left(t_{\text {cancer }} \leq k \mid \widetilde{R}\right) =B(\widetilde{R})+\sum_{i=0}^{k-1} H_i(\widetilde{R})
\end{equation}

which facilitates end-to-end optimization via log-likelihood maximization on the observed screening follow-up data while ensuring temporal consistency in risk estimation.

\section{Experiments}
\subsection{Dataset}
Following LoMAR~\cite{LoMaR}, this study uses the publicly available CSAW-CC (mammography) dataset~\cite{CSAW-CC} to ensure reproducibility. In total, the dataset includes 873 breast cancer cases out of 1,103 initially identified patients, and 7,850 healthy controls out of 10,000 randomly selected individuals. For each subject, all screening mammograms from multiple time points are provided, along with the dates of examination and patient age. Additionally, pixel-level tumor annotations are included for cases, indicating the precise region of the tumor. We adhere to the preprocessing protocol outlined in LoMaR~\cite{LoMaR}; detailed information on class distribution can be found therein, as the dataset utilized in the current study closely aligns with that work due to the identical preprocessing strategy.

\subsection{Experimental details}
We initialized the VMRNN blocks with the publicly released weights from \cite{VMRNN} and initialized the image encoder with Mirai~\cite{Mirai} weights. Additionally, the image embeddings used by the VMRNN and asymmetry block are obtained from Mirai’s image encoder, which remains frozen during training. We employ AdamW with an initial learning rate of $1 e-3$, weight decay of $1 e-4$, and a cosine decay schedule. The number of training epochs is 30. To handle class imbalance and improve risk prediction accuracy, we use a weighted cross-entropy loss with class weights derived from the training set distribution, following LoMaR\cite{LoMaR}. Following, conventional risk prediction assessment~\cite{Mirai,LoMaR}, we utilized the C-index and area under the receiver operating characteristic (ROCAUC) curve to evaluate the performance of all methods in the 1–5 year risk prediction task. All experiments were conducted using 8× NVIDIA A100 GPUs with 80GB RAM.

\subsection{Results}
The performance of risk prediction in various models is shown in Table~\ref{tab:Res}. Notably, OncoNet, Mirai, and AsymMirai do not incorporate historical data, whereas the other models leverage four years of such data to provide a richer temporal context. Our approach emphasizes two primary components: the VMRNN module for robust temporal encoding and the asymmetry module for capturing spatial discrepancies over time. For clarity, these components are denoted as “VMRNN” (VMR) and “Asymmetry” (Asym).

The results demonstrate that compared to the current SOTA model LoMaR, our VMRA-MaR model which integrates both the VMR and Asym modules, shows notable improvements, particularly at the 4 and 5 year follow-up intervals (C index of 0.82 with ROCAUC scores of 0.84 at both time points). Although the p-value of 0.061 suggests only a borderline significant difference from LoMaR, the trend indicates potential gains from combining temporal encoding with asymmetry features. Further analyses among VMRNN-based models suggest that while incorporating asymmetry can improve performance—especially at earlier follow-up intervals—the magnitude of these benefits may vary over longer horizons. The superior performance of VMRNN is attributed to its dynamic recurrent updating, gated mechanisms, and integrated spatial feature extraction, which allow it to effectively capture long-range dependencies and complex temporal relationships—providing a significant advantage over Transformer models that rely on fixed positional embeddings.

Note, while our design for incorporating the asymmetry score averages these scores, we have further experimented with alternative approaches such as concatenating all asymmetry scores across time steps. Our comparisons demonstrated that the average fusion consistently yielded the best performance, reinforcing the effectiveness of this approach.

\begin{table}[!t]
    \centering
    \scriptsize
    \caption{Ablation results on breast cancer 5-year risk prediction with C-index and ROCAUC scores. * indicates that the model uses historical data. VMR: Vision-Mamba RNN; Asym: Asymmetry-Aware Risk Factor Extraction.}
    \begin{adjustbox}{center}
    \begin{tabular}{c|c|c|c|*{5}{c}}
    \toprule
    \multirow{2}{*}{Model} & 
    \multicolumn{2}{c|}{Module} & 
    \multirow{2}{*}{C-index} & 
    \multicolumn{5}{c}{Follow-up year ROCAUC} \\
    \cline{2-3} \cline{5-9}
    & VMR & Asym & & 1-year & 2-year & 3-year & 4-year & 5-year \\
    \hline
    \makecell{OncoNet\\(Only-DL)}  & \ding{55} & \ding{55} & \makecell{0.75\\(0.72-0.78)} & \makecell{0.82\\(0.80-0.84)} & \makecell{0.80\\(0.78-0.82)} & \makecell{0.75\\(0.73-0.78)} & \makecell{0.73\\(0.71-0.75)} & \makecell{0.68\\(0.67-0.69)} \\
    \hline
    Mirai & \ding{55} & \ding{55} & \makecell{0.82\\(0.78-0.83)} & \makecell{0.92\\(0.90-0.94)} & \makecell{0.84\\(0.81-0.87)} & \makecell{0.80\\(0.78-0.82)} & \makecell{0.79\\(0.78-0.80)} & \makecell{0.77\\(0.76-0.79)} \\
    \hline
    AsymMirai & \ding{55} & \checkmark & \makecell{0.81\\(0.78-0.83)} & \makecell{0.80\\(0.74-0.86)} & \makecell{0.68\\(0.64-0.71)} & \makecell{0.68\\(0.66-0.70)} & \makecell{0.67\\(0.65-0.69)} & \makecell{0.64\\(0.60-0.68)} \\
    \toprule
    LoMaR* & \ding{55} & \ding{55} & \makecell{0.82\\(0.78-0.85)} & \makecell{0.92\\(0.89-0.95)} & \makecell{0.83\\(0.85-0.86)} & \makecell{0.82\\(0.84-0.86)} & \makecell{0.80\\(0.78-0.82)} & \makecell{0.83\\(0.85-0.87)} \\
    \hline
    VMR\_MaR* & \checkmark & \ding{55} & \makecell{\textbf{0.84}\\(0.79-0.89)} & \makecell{0.91\\(0.90-0.93)} & \makecell{0.82\\(0.80-0.84)} & \makecell{\textbf{0.84}\\(0.82-0.86)} & \makecell{0.82\\(0.80-0.84)} & \makecell{\textbf{0.86}\\(0.84-0.86)} \\
    \hline
    VMRA\_MaR* & \checkmark & \checkmark & \makecell{0.82\\(0.80-0.83)} & \makecell{\textbf{0.94}\\(0.93-0.95)} & \makecell{\textbf{0.88}\\(0.86-0.89)} & \makecell{0.82\\(0.79-0.84)} & \makecell{\textbf{0.84}\\(0.82-0.86)} & \makecell{0.84\\(0.82-0.86)} \\
    \toprule
    \end{tabular}
    \end{adjustbox}

    \label{tab:Res}
\end{table}

\begin{table}[t]
\scriptsize
\centering
\caption{Subgroup results on risk detection across low, medium, and high breast density were combined with follow-up year ROCAUC.}
\resizebox{\textwidth}{!}{%
\begin{tabular}{c|c|c|c|c|c|c|c|c|c|c|c|c|c|c|c}
   \toprule
   \multirow{2}{*}{\textbf{Model}} 
   & \multicolumn{3}{c|}{\textbf{1-year}} 
   & \multicolumn{3}{c|}{\textbf{2-year}}
   & \multicolumn{3}{c|}{\textbf{3-year}} 
   & \multicolumn{3}{c|}{\textbf{4-year}}
   & \multicolumn{3}{c}{\textbf{5-year}} \\  
   \cline{2-16}
   & Low & Med & \textbf{High}
   & Low & Med & \textbf{High}
   & Low & Med & \textbf{High}
   & Low & Med & \textbf{High}
   & Low & Med & \textbf{High} \\
   \hline
   OncoNet & 0.85 & 0.82 & 0.79 & 0.88 & 0.78 & 0.70 
             & 0.76 & 0.76 & 0.73 & 0.73 & 0.75 & 0.71 
             & 0.72 & 0.74 & 0.58 \\
   Mirai & 0.94 & \textbf{0.95} & 0.87 & 0.80 & 0.86 & 0.86
             & 0.81 & 0.80 & 0.79 & 0.81 & 0.77 & 0.79
             & 0.78 & 0.76 & 0.77 \\
   AsymMirai & 0.84 & 0.86 & 0.70 & 0.72 & 0.66 & 0.66
             & 0.73 & 0.68 & 0.63 & 0.68 & 0.65 & 0.66
             & 0.70 & 0.62 & 0.60 \\
   LoMaR & \textbf{0.95} & 0.93 & 0.88 & \textbf{0.92} & 0.82 & 0.75
             & \textbf{0.89} & 0.82 & 0.75 & 0.84 & 0.81 & 0.75
             & \textbf{0.87} & 0.83 & 0.79 \\
    \hline
   VMR\_MaR & 0.93 & 0.89 & 0.90 & 0.90 & 0.80 & 0.84
             & 0.83 & \textbf{0.84} & \textbf{0.85} & 0.82 & 0.80 & \textbf{0.84}
             & 0.80 & \textbf{0.91} & \textbf{0.87} \\
   VMRA\_MaR & 0.92 & \textbf{0.95} & \textbf{0.97} & 0.85 & \textbf{0.89} & \textbf{0.90}
             & 0.85 & 0.78 & \textbf{0.85} & \textbf{0.86} & \textbf{0.82} & \textbf{0.84}
             & 0.80 & 0.87 & 0.85 \\
   \bottomrule
\end{tabular}
}
\label{tab:Density}
\end{table}

\begin{figure}[t]
    \centering
    \includegraphics[width=\textwidth]{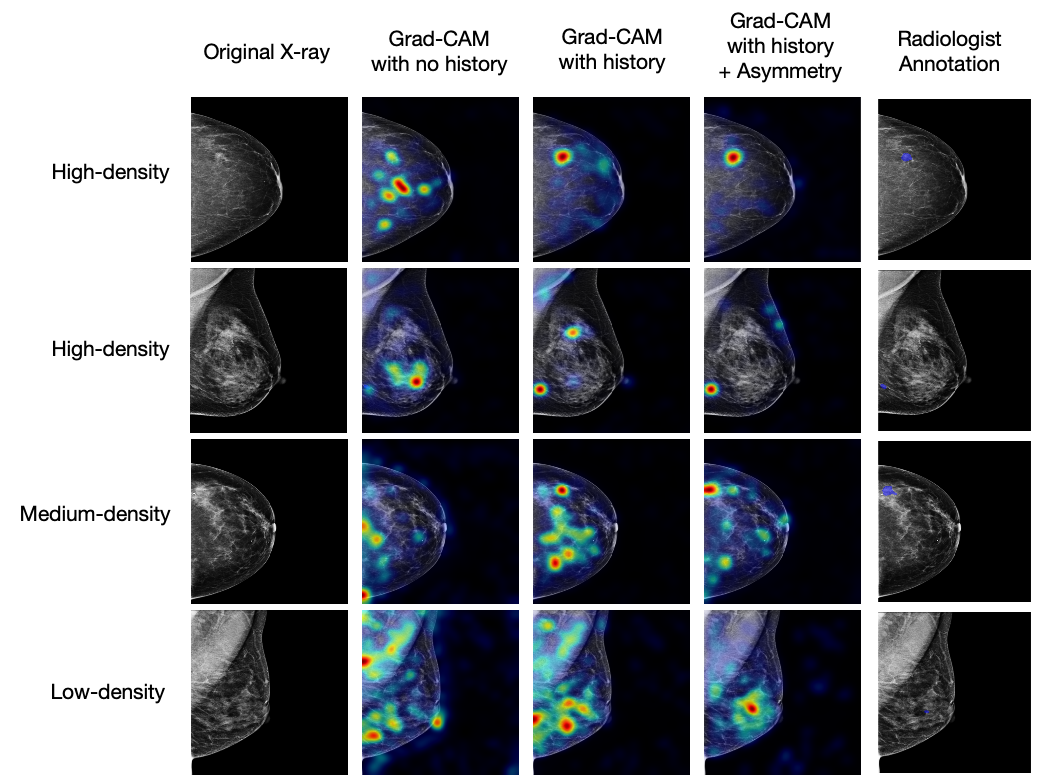}
    \caption{Grad-CAM visualizations of VMR-MaR and VMRA-MaR models for five representative subjects (two high-, one medium-, and one low-density).}
    \label{fig:Grad}
\end{figure}

\subsection{Subgroup Analysis on Risk Detection Across Breast Density}
To further provide insights into the benefits of leveraging temporal information compared to the current state-of-the-art model LoMaR, we analyzed performance stratified according to breast density (see Table~\ref{tab:Density}). In particular, we categorized the data into three breast density levels (low, medium, and high) based on the metric \texttt{libra\_densearea} in the CSAW-CCS dataset\cite{CSAW-CC}. We utilize this absolute quantity of dense breast tissue \texttt{libra\_densearea} as it has been shown that absolute dense tissue tends to provide more relevant information on breast cancer risk compared to percentage density (see for instance\cite{DA}). This metric, which represents the dense area in cm\textsuperscript{2} estimated by the Libra software, was first sorted for all patients and then trichotomized to define the low, medium, and high subgroups. This stratification is especially relevant because higher breast density not only increases the intrinsic risk of developing breast cancer but also complicates the detection of lesions on mammograms.

Notably, our VMRA-MaR model demonstrates a clear advantage in the high-density subgroup: In the 1-year follow-up, it achieves a ROCAUC of 0.97 compared to 0.88 for LoMaR, 0.90 for VMR\_MaR, and lower scores for OncoNet (0.79)\cite{OncoNet,OncoNet1,OncoNet2} and Mirai (0.87). This superior performance persists at 2 years (ROCAUC of 0.90) and remains consistent across the 3-, 4-, and 5-year intervals, underscoring its robust ability to extract and utilize complex features from dense mammary tissues.

We hypothesize that while models lacking temporal analysis perform adequately in low density cases, where tissue characteristics are less challenging the incorporation of temporal features is particularly beneficial for detecting tumors in high-density breast tissue. Our asymmetry-aware, recurrent temporal framework effectively captures intricate tissue dynamics, thereby enhancing performance in this subgroup.

\subsection{Exploration of model focus by Grad-cam}
Figure \ref{fig:Grad} presents Grad-cam \cite{grad-cam} saliency maps derived from mammographic X-ray images, with the rightmost column displaying expert annotations of cancer lesions. Comparative analysis indicates that incorporating temporal information significantly improves lesion localization, while integrating asymmetry features further refines its precision.

\section{Conclusion}
In this study, we introduce an Asymmetry-Aware Temporal Framework for breast cancer risk prediction, VMRA-MaR. Our unified approach combines a VMRNN architecture with asymmetry modules to dynamically capture evolving spatial cues and long-range temporal dependencies, yielding accurate risk estimates. The results indicate that our framework surpasses the previous state-of-the-art, LoMaR, by achieving higher accuracy at extended time steps and improved risk prediction for the challenging high-density breasts. Future work will focus on refining the alignment of mammography views to further enhance lesion localization and overall predictive precision.

\begin{credits}
\subsubsection{\ackname} 
This work was supported by The Research Council of Norway (Visual Intelligence, grant no. 309439 as well as FRIPRO grant no. 315029 and IKTPLUSS grant no. 303514).

\subsubsection{\discintname}
The authors have no competing interests to declare that are relevant to the content of this article.
\end{credits}

%
%
%

\bibliographystyle{splncs04}
\bibliography{VMRA_MaR/main}
\end{document}